\documentclass{frontiersSCNS} 

\usepackage{url,lineno}
\usepackage{color}
\usepackage{todo}

\def\keyFont{\fontsize{8}{11}\helveticabold }
\def\firstAuthorLast{Sagarra {et~al.}} 
\def\Authors{Oleguer Sagarra\,$^{2}$, Mario Guti\'errez-Roig\,$^{1,2}$, Isabelle Bonhoure\,$^{1}$, and Josep Perell\'o\,$^{1,2,*}$}
\def\Address{$^{1}$OpenSystems Research, Departament de F\'isica Fonamental, Universitat de Barcelona, Barcelona, Spain \\
$^{2}$Complex Lab Barcelona, Departament de F\'isica Fonamental, Universitat de Barcelona, Barcelona, Spain}
\def\corrAuthor{Josep Perell\'o}
\def\corrAddress{Departament de F\'isica Fonamental, Universitat de Barcelona, Mart\'i i Franqu\'es 1, Barcelona, E-08028, Spain}
\def\corrEmail{josep.perello@ub.edu}

\begin{document}
\onecolumn
\firstpage{1}

\title[Citizen Science practices]{Citizen Science practices for Computational Social Sciences research: The conceptualization of Pop-Up Experiments}
\author[\firstAuthorLast ]{\Authors}
\address{}
\correspondance{}
\extraAuth{}
\topic{}

\maketitle


\begin{abstract}
Under the name of Citizen Science, many innovative practices in which volunteers partner with scientist to pose and answer real-world questions are quickly growing worldwide. Citizen Science can furnish ready made solutions with the active role of citizens.
However, this framework is still far from being well stablished to become a standard tool for Computational Social Sciences research. We present our experience in bridging Computational Social Sciences with Citizen Science philosophy, which in our case has taken the form of what we call Pop-Up Experiments: Non-permanent, highly participatory collective experiments which blend features developed by Big Data methodologies and Behavioural Experiments protocols with ideals of Citizen Science. The main issues to take into account whenever planning experiments of this type are classified and discused grouped in three categories: public engagement, light infrastructure and knowledge return to citizens. We explain the solutions implemented providing practical examples grounded in our own experience in urban contexts (Barcelona, Spain). We hope that this work serves as guideline to groups willing to adopt and expand such \emph{in-vivo} practices and opens the debate about the possibilities (but also the limitations) that the Citizen Science framework can offer to study social phenomena.

\tiny
 \keyFont{ \section{Keywords: Citizen Science, Participation, Engagement, Computational Social Science, Data, Experiments, Collective, Methods} } 
\end{abstract}

\section{Introduction}

The relationship between science and society has always been an important aspect to consider when one tries to understand how science advances and how research is performed~\citep{latour:1998,kuhn:2012}. The general public has however been mostly left out of its methodology and creation processes~\citep{latour:2006,callon:2003}. Citizens are generally considered as passive subjects to whom only finished results can be presented in the form of simplified statements, yet paradoxically, we implicitly ask them to support and encourage research. The acknowledgement of this \textit{Ivory Tower} problem has recently opened new and exciting opportunities to open minded scientists. The advent of digital communication technologies, mobile devices and the Web 2.0 is fostering a new kind of relation between professional scientists and dedicated volunteers or participants. 

Under the name of Citizen Science (CS), many innovative practices in which ``volunteers partner with scientist to answer and pose real-world questions" (as stated in the Cornell Ornithology Lab webpage, one of the precursors of CS practices in 1980s) are quickly growing worldwide~\citep{hand:2010,bonney:2014,gura:2013,editorial:2015}. Formally, CS has been recently defined by the Socientize White Paper by a ``general public engagement in scientific research activities when citizens actively contribute to science either with their intellectual effort or surrounding knowledge or with their tools and resources"~\citep{whitepaper:2014}. This open, networked and transdisciplinary scenario, favours a more democratic research thanks to amateur or non professional scientists' contribution~\citep{irwin:1995}. During last few years, relevant results have been published in high impact journals such as Nature or Science by using participatory practices~\citep{bonney:2014,gura:2013}. The hidden power of thousands of hands working together is making itself apparent in many fields, with comparable (or even better) performance to expensive supercomputers when used to analyse/classify astronomical images~\citep{lintott:2008}, to reconstruct 3D brain maps based on 2D images~\citep{Kim:2014}, or to find out stable biomolecular structures~\citep{cooper:2010} just to name very few large impact cases. Citizen contributions can also have a direct impact in society by for instance helping to create exhaustive and shared geo-localised datasets~\citep{lauro:2014} at a density level unattainable for the vast majority of private sensor networks (and at a much reduced cost) or by collectively gathering empirical evidences to force public administration action (for example the shutdown of a noisy factory located in the city of London~\citep{haklay:2013}). Most active volunteers can therefore contribute by providing experimental data and widen the researchers reach, raise new questions and co-create a new scientific culture~\citep{latour:2006,mcquillan:2014}.

Computational Social Sciences (CSS) is a multidisciplinary field at the intersection of the social, computational and complexity sciences which subject of study is the human interactions and society itself ~\citep{lazer:2009,cioffi:2010}. However, CS practices in this context remain vastly unexplored when compared to others fields such as Environmental Sciences which already carry a long story on their shoulders \citep{russell:2014,dickinson:2015,silvertown:2009}. Attempts where participation of ordinary citizens has played an important role can be found within fields like Experimental Economics~\citep{hoggatt:1959}, financial trading floors design~\citep{cueva:2015}, or human mobility~\citep{yoshimura:2014}. Also works on the emergence of cooperation~\citep{grujic:2010} and the dynamics of social interactions~\citep{weiss:2014} are noteworthy. 
All these experiments yielded relevant scientific outcomes with well-stablished and robust protocols within Behavioural Sciences (see for instance \cite{kagel:1997} and \citep{margolin:2013}) but unfortunately they stay on the very first level of CS scale~\citep{haklay:2013}, where citizens are involved only as sensors or volunteer subjects of certain experiment in strictly controlled environments, being their participation and potential only partially unleashed. A possible way out was already provided by~\cite{latour:2006,latour:2009} when proposing collective experiments in which the public becomes a driving force
of the research. Researchers \emph{in the wild} are then directly concerned with the knowledge they produce
because they are both the objects and the subjects of their research \citep{callon:2003}. Some interesting research initiatives have emerged along these lines to run massive experiments in collaboration with a CS foundation such as Ibercivis \citep{GraciaLazaro:2012} or through online platforms such as \emph{Volunteer Science} online platform from Lazer Lab. More radical initiatives also consider the collaboration with artists and have been carried out in museums or exhibitions and as big performances \citep{perello:2012,cattuto:2010}.

CSS research has also been recently working within the so called Big Data paradigm \citep{schroeder:2014,conte:2012}. A lot has been said about it and the possibilities it offers to society, industry and researchers. Smart Cities densify urban areas with all kind of sensors and integrate the information with a wide collection of datasets. Mobile devices represent also a powerful tool to monitor on real time user related stats such as health and great businesses opportunities are already being foreseen by companies. However, these approaches again treat citizens as passive subjects from whom one records private data in an unconsented way, with the aggravated problem that the \emph{unaware} producers of this data (i.e., citizens) lose control of its use, exploitation and analysis. Concerning this last aspect, the validity of the conclusions drawn from the analysis of such datasets are still today a subject of discussion, mainly due to the poor control on the data gathering process (by the public in general and by scientists in particular), inherent population and sampling biases \citep{Lazer:2014} and lack of reproducibility among other systemic problems \citep{Boyd:2012}. Last but not least, the Big Data paradigm has so far failed in providing society with the necessary public debate and transparent practices from the \emph{bottom-up} approach it preconizes: It currently relies on huge infrastructures only available to private corporations, which may not have coincident objectives with those of researchers and citizenry; and provide conditioned access to the data contents which, in addition generally cannot be freely (re)used without filter.

Our purpose here is not to discuss some of the Big Data inherent problems. The approach we present aims however to explore the potential in blending interesting features recently developed by Big Data methodologies with the ambitious and democratic ideals of CS. Public participation and scientific empowerment induce a level of (conscious) proximity with the subjects of the experiments that can be a very valuable source of high quality data \citep{vianna:2014,rossiter:2015}, or at least, of non-conflictive information with regards to data anonymity \citep{Bowser:2014}, that may be considered to correct biases and experimental systematic errors. This approach is a potential way to overcome privacy and ethical issues arisen when collecting data from digital social platforms while keeping high standards of participation~\citep{schroeder:2014,resnik:2015,riesch:2013}. Moreover, CS projects use a vast variety of social platforms for optimizing dissemination, to encourage and increase participation and sometimes to develop \textit{gamification} strategies \citep{prestopnik:2011} to reinforce such engagement. The so-called Science of Citizen Science studies the emergent participatory dynamics in this class of projects~\citep{sauermann:2015,curtis:2015} so that this also opens the door of new contexts to study social phenomena. 

The open philosophy at the heart of CS methods, like open-data licensing and coding, can also clearly improve science-society-policy interactions in a democratic and transparent way \citep{petersen:1984}. The CS approach simultaneously represents a powerful example of Responsible Research and Innovation (RRI) practices included in the EU Horizon 2020 research programme~\citep{eu:2015} and the Quadruple Helix Model where government, industry, academia and civil participants work together to co-create the future and drive structural changes far beyond the scope of what any one organization or person could do alone~\citep{chesbrough:2006}. Along this frame, we consider that the CSS potential when assuming CS methods is vast since its subject of study are the citizens themselves. Therefore, their engagement with projects studying their own behaviour is highly likely since it has an immediate impact in their daily life. As a result, large motivated communities and scientists can work hand in hand to tackle CSS' arising challenges but also collectively circumvent potential side effects. The possibility to reach wider and more diverse communities will help in refinement of more universal statements avoiding population biases~\citep{henrich:2010} and problems of reproducibility present in Social Sciences empirical studies \citep{Open:2015,Bohannon:2015}. Another important advantage of working jointly with different communities is that it allows scientists to set \emph{lab-in-the-field} or \emph{in-vivo} experiments instead of isolating subjects from their natural urban environment --where socialization takes place-- but do it in a transparent, fully consented and enriched way thanks to the active participation of citizens \citep{callon:2003,perello:2012}. Such practices and methodologies, however, are still far from being well stablished to become a standard tool for CSS research.

The main goal of this paper is precisely to motivate the somewhat unexplored adoption of CS practices by CSS research activities. Along this arduous task, we limit ourselves to a {\it reformatting} of existing standard experimental strategies and methods in science through what we call Pop-Up Experiment (PUE). Such a concept has been shaped from our own lessons gained while running experiments in public spaces of the city of Barcelona (Spain). Section \ref{Sect2} introduces this very flexible solution to make possible a collective experimentation and discusses its three essential ingredients: public engagement, adaptable infrastructure, and knowledge return to citizens. Finally, Sect. \ref{Discussion} concludes the manuscript with a discussion of what we have presented so far jointly with some considerations about the future of CS practices within the CSS research.

\section{A flexible solution for Citizen Science practices inside CSS: The Pop-Up Experiments}
\label{Sect2}

\subsection{Context and motivation}

During the last four years, the Barcelona Council and its Direction of Creativity and Innovation in collaboration with several organisations have aimed to explore the possibilities to transform the city into a public \emph{living lab}~\citep{chesbrough:2006} where new creative technologies can be tested and new knowledge can be collectively constructed. This has been done through the Barcelona Lab platform and one of its most preeminent actions has been to establish CS practices in and with the city. The first task led by some of us has been to create the Barcelona Citizen Science Office and to build a community of practitioners where most of the CS projects from several research institutions in Barcelona could converge. The Office serves as \emph{meeting point} for CS projects, where researchers can join forces, experiences and knowledge and also where citizens can connect with these initiatives in an easy and effective way. The second task is directly linked with the object of this paper and was conceived to test how far the public administration can go in opening up their resources to collectively run scientific experiments \citep{perello:2012}. The CS toolbox clearly provides a perfect framework for the design of public experiments and the exploration of the emergent tensions and the problematic issues faced when running public living labs. Furthermore, the involvement of the City Council provided us with the opportunity to embed these experiments into important and massive cultural events, which constitute a perfect environment to reinforce the openness and the transparency in our research process in front of the society or at least in front of the Barcelona citizens.

\begin{table}[!t]
\textbf{\refstepcounter{table}\label{Tab:exps} Table \arabic{table}. }{Summary of our seven experiments to test Citizen Science practices with Pop-Up Experiments in urban contexts and along the Computational Social Sciences research. First column provides the different aspects that each of the experiments has considered in terms of Experimental Infrastructure and Set-Up, Public Engagement tools and Outcome and Return to the Public. The other columns specify how the different experiments consider these aspects. Further details are provided in Material and Methods section while Sect. \ref{Sect2} deeper discuss the different aspects.}
\includegraphics[width=18.7cm]{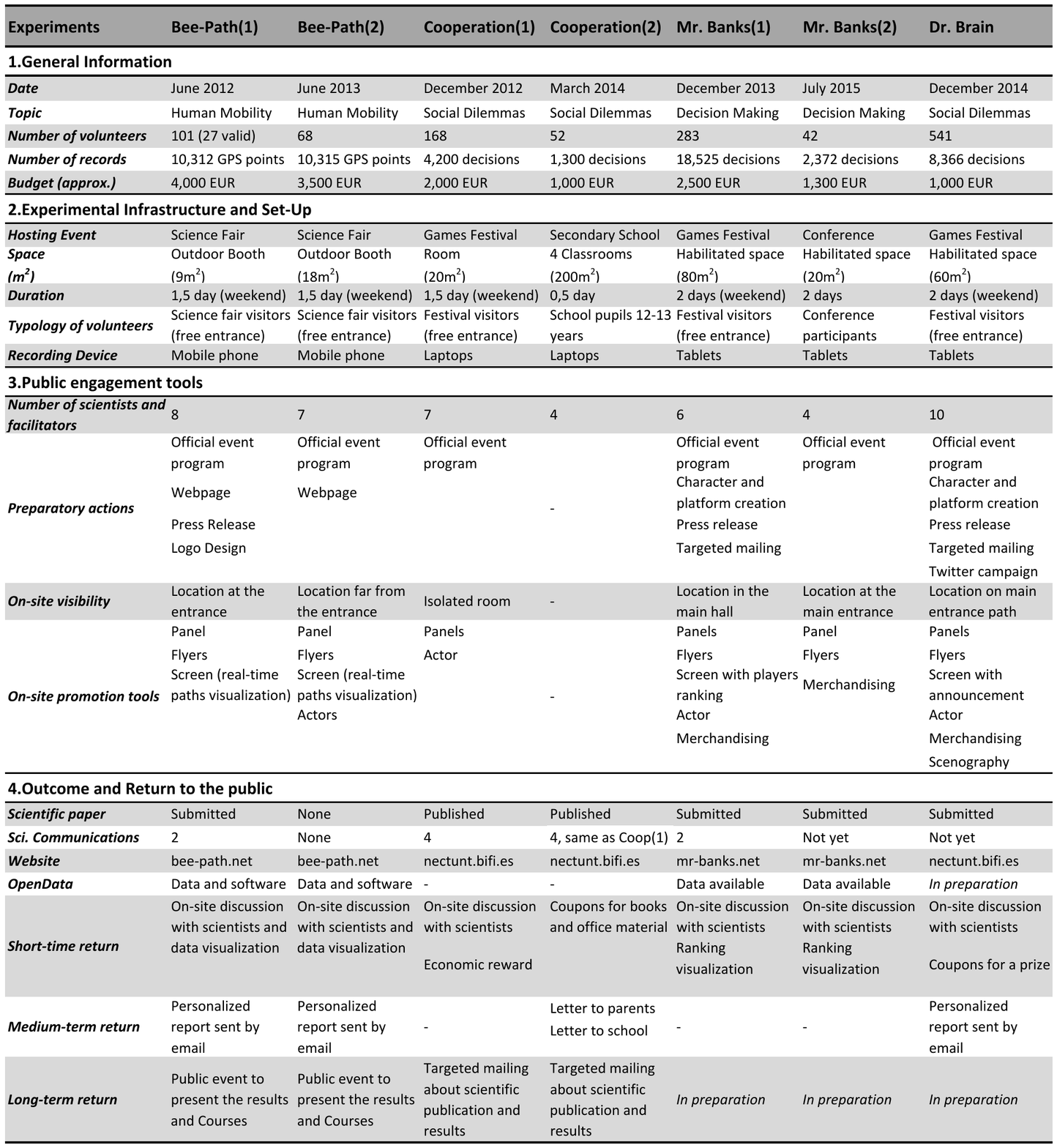}
\end{table}

We have been conducting several experiments to bring CS ideals into practice and test its potential in urban contexts. In contrast with existing environmental CS projects in other cities such as London or New York (with civic initiatives such as \emph{Mapping for Change} or \emph{Public Lab}, respectively) we focussed our attention to CSS related problems with the aim of providing unexplored relations between city, citizens and scientists. More specifically, this rather wild testing comprised seven different experiments performed between 2012 and 2015 in three different topics addressing diverse questions: human mobility (How do we move?), social dilemmas (How cooperative are we?) and decision-making process (How do we take decisions in a very uncertain environment like financial markets?). These are summarized in Tab. \ref{Tab:exps} and fully described in Section \ref{sec:methods} (Material and Methods). Other common points of these experiments are the important number of volunteers participating (up to 541 for a single experiment, 1,255 in total) and the consequent large number of records (up to 18,525 decisions for a single experiment, 55,390 records in total), despite a rather limited budget associated to the experiments (from 1,000 to 4,000 EUR/experiment, and around 2,200 EUR/experiment on average). Despite dealing with different research questions, common problems (related to CS practices but also CSS research studying human behaviour~\citep{conte:2012}) have been identified and solutions have been developed in all cases to overcome them. Some of the implemented solutions have been successful some others not, but all experiences have shaped the concept and the process for an experimentation in CSS research consistent with CS ideals. 

\subsection{Pop-Up Experiment definition and underlying process}

The generic definition of a \emph{pop-up} according to the Cambridge dictionary is ``Pop-up (adj.): used to described a shop, restaurant, etc. that operated temporarily and only for a short period when it is likely to get a lot of customers." From early beginning we thought that this description fully fits into our non-permanent but highly participatory experimental set up when applying CS principles to Computational Social Sciences research in urban contexts. The parallelism is very illustrative to better understand a much more formal definition we use to describe our approach from a theoretical perspective. It is based on the expertise gained from the seven experiments carried out during the last four years and reads: 
\begin{quotation}
A Pop-Up Experiment (PUE) is a physical, light, very flexible, highly adaptable, reproducible, transportable, tuneable, collective, participatory and public experimental set-up for urban contexts that (1) applies Citizen Science practices and ideals to provide ground-breaking knowledge and (2) transforms the experiment into a valuable, socially responsible, consented and transparent experience to non-expert volunteered participants with the possibility to build the urban commons arisen from facts-based effective knowledge valid for both cities and citizens.
\end{quotation}

\begin{figure}[t]
\begin{center}
\includegraphics[width=18cm]{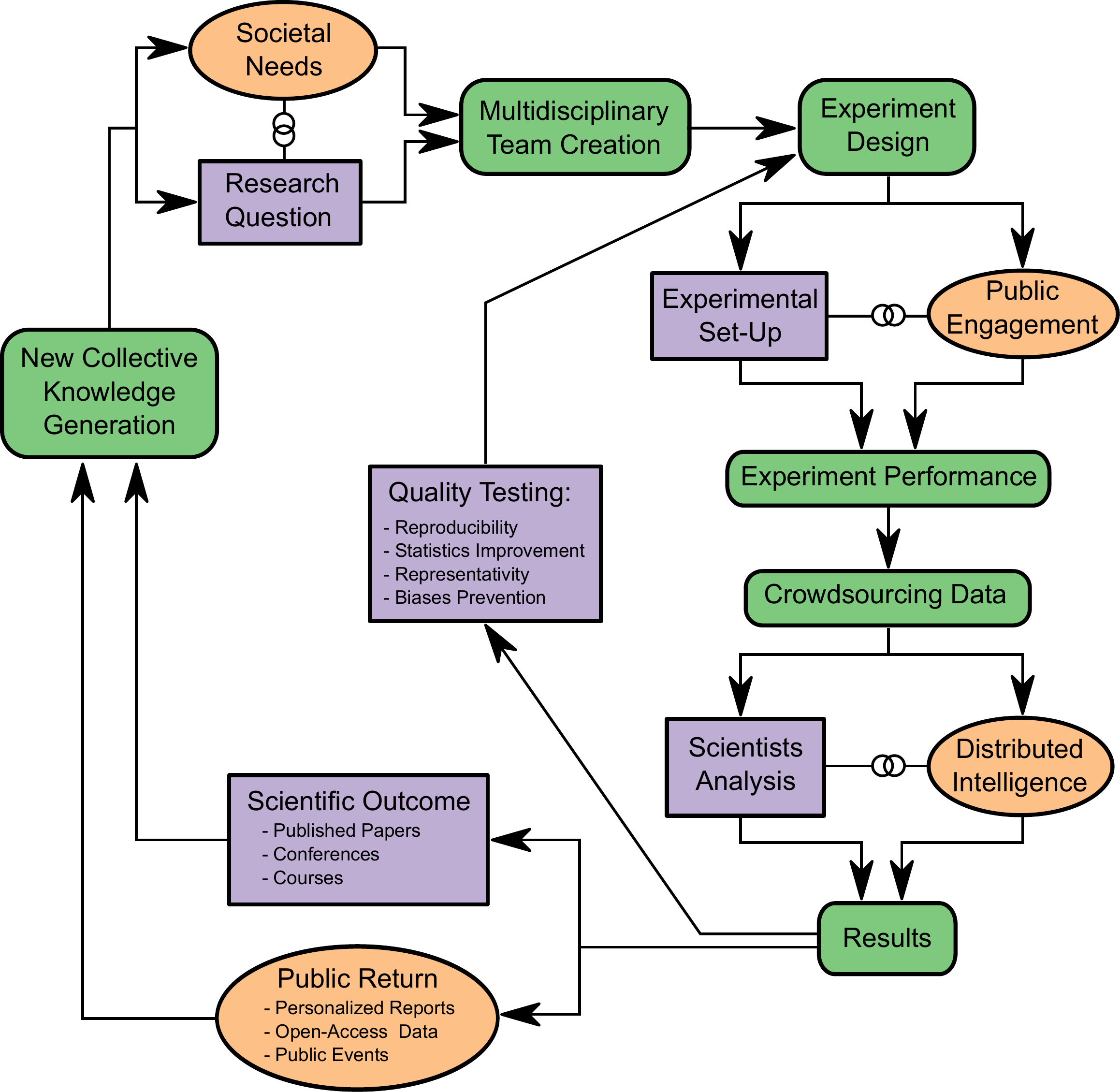}
\end{center}
\textbf{\refstepcounter{figure}\label{fig:01} Figure \arabic{figure}. }{Research process in Pop-up Experiments. We identify the different steps of the whole process in a set of boxes, from its conception to its completion. Starting from the research question, we introduce the design and the performance of the given experiment aiming to respond to a particular challenge. Crowsourced data gathered can be analyzed both by expert scientists and citizen amateurs to produce new knowledge (return) in multiple forms. The results are finally taken as inspiration and renewed energy to face new challenges and new research questions. Data quality is also part of the process: lessons learnt from the scientific analysis can improve future experimental conditions. Oval forms in orange correspond to volunteers' contributions, squared forms in magenta are tasks exclusively done by scientists while rounded rectangles in green are shared tasks by both citizens and scientists.}
\end{figure}

In our case, we apply this concept to CSS aiming to answer very concrete research questions with the participation of larger populations than those from Behavioural Experiments. The arising research process from PUEs can be synthesized in the flow diagram from Fig. \ref{fig:01}. 
The whole process starts from a research question and/or a societal challenge that may be promoted by citizens, scientists but also private organisations, public institutions or civic movements. The initial impulse helps to create an adequate research group which will need to be multidisciplinary if it is to tackle a complex problem with many interwined issues. The group then co-creates the experiment both considering the experimental set-up and the unavoidable public engagement tasks. The experiment is then carried out and data is 
collectively generated (crowsourced) under the particular constrains of public spaces which not only depend on the conditions designed by the scientists but also on many other practical limitations. The data is afterwards analyzed using standard scientific methods but non-professional scientists are also invited to contribute on specific tasks \citep{Kim:2014,lintott:2008} or by using other non-standard strategies in the data exploration \citep{cooper:2010}. These two contributions by volunteers conform what we call Distributed Intelligence and generate results that can hardly be matched by conventional computer analysis.
In any case, obtained results can take many forms according to the audience being addressed, from a scientific paper to personalized reports readable to any citizen or even recommendations valid for policymakers at a city level. Finally, the whole process can generate the necessary inertia to promote and face a new societal and scientific challenge or an existing need along the same scheme. 

\begin{figure}[t]
\begin{center}
\includegraphics[width=18cm]{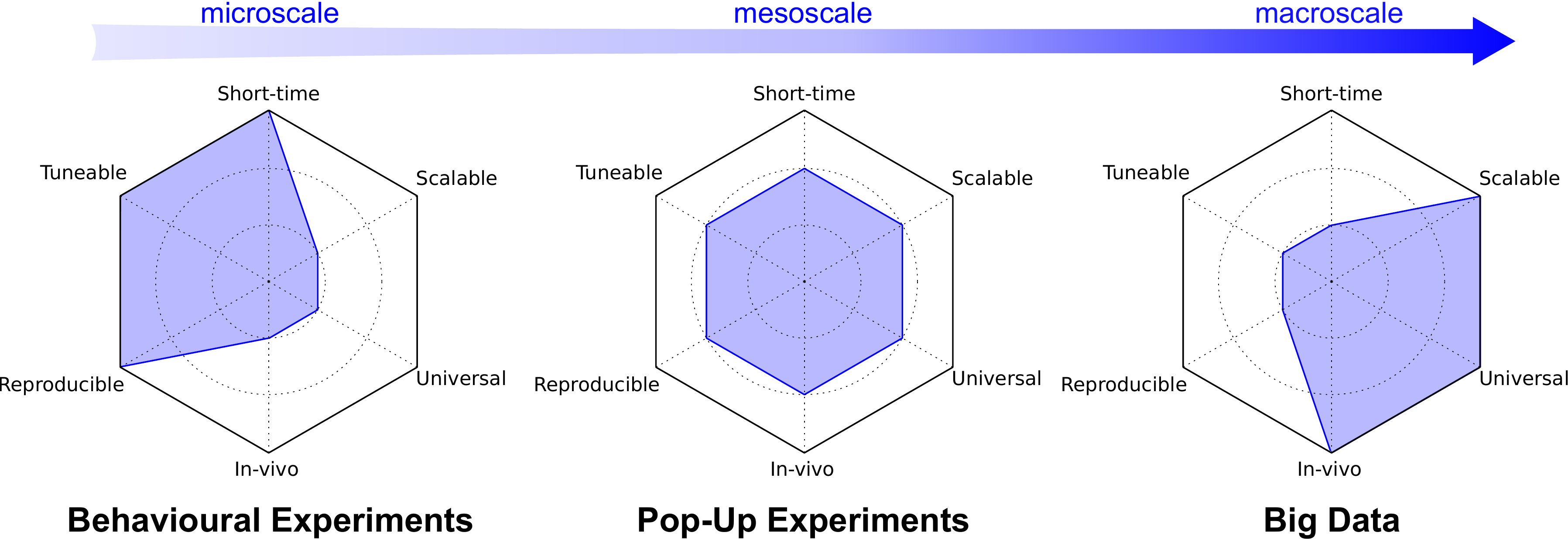}
\end{center}
\textbf{\refstepcounter{figure}\label{fig:02} Figure \arabic{figure}. }{Radar Chart comparing seven characteristics of the Pop-Up Experiment approach with the research made through Behavioural Experiments and Big Data. Features are graded in three different degrees of intensity from low (shortest radius) to high (largest radius). "Short-time" describes the time required for running the experiment. "Scalable" qualifies how easy is to upscale and augment the number of subjects in the experiment while preserving original design. "Universal" quantifies the generality of the statements produced by the experiments. "In-vivo" measures how close is the experimental set-up to quotidian situations or everyday life situations. "Reproducible" assesses the capacity to repeat same experiment under identical conditions. Finally, "Tunable" quantifies how flexible and versatile are the conditions of the experimental design.}
\end{figure}

The PUE solution also represents a middle ground between Behavioural Science experiments and Big Data methodologies.
To better understand in which context the PUEs we propose can be placed, Fig. \ref{fig:02} compares the different approaches considered using a radar chart that qualitatively measures with three degrees of intensity (low, medium and high) six different aspects characterizing each experiment type. We observe that the three different approaches cover different areas. Behavioural Experiments and Big Data have a limited overlap while the PUEs share several aspects with the former two. One might argue that the \emph{excess} of openness of the CS constitutes a severe limitation to raise objectivity in comparison to the solid experimental protocols in Behavioural Sciences \citep{webster:2014,kagel:1997} but it is also true that the highly participatory nature of CS can be very effective to reach a more realistic spectra of population and a larger sample thus obtaining more general statements with a stronger statistical support (see \cite{margolin:2013} for alternative and complementary methods). Since it is directly attached to real-world situations, it avoids the somehow exclusive and sometimes distorted space of the \emph{in-vitro} (or \emph{ex-vivo}) laboratory experiences. It also brings additional values to the more classic lab-in-the-field experiences in Social Sciences which generally limit interaction among subjects and scientists as much as possible. On the other extreme, CS practices will never be able to compete in terms of data quantity 
with the Big Data world but this can be compensated. A better understanding of the mass of volunteers involved and improved knowledge of its peculiarities helps in avoiding possible biases. Furthermore, the active nature of PUEs allow the possibility of tunning some conditions of the experiments to explore alternative scenarios. The PUEs can indeed be an alternative to the controversial virtual labs in social networks and mobile games which have obtained for instance interesting results on emotional contagion with experiments on the Facebook platform~\citep{Kramer:2014} not without an intense public debate on ethical and privacy issues on the way how the experiments were made \citep{gleibs:2014}.

We think that PUEs can be an essential approach for empirically testing the abundant statements of CSS complementary to lab-in-the-field, virtual labs and in-vitro experiences. To become so, we have identified the main obstacles that hinder the development of CS initiatives with respect to other forms of social experimentation. They can be grouped into three groups: Infrastructure, Engagement, and Return. In the following, we detail each of the obstacles and illustrate the solutions that PUEs offer together with practical examples applied to each case.


\subsection{Light and flexible experimental infrastructure}

By infrastructure we understand all necessary logistics to make the experiments possible. In a broad Citizen Science context, the necessary elements differ from orthodox scientific infrastructure. As discussed in \cite{bonney:2014} and \cite{franzoni:2014}, they include other tools, other technical support and other spaces. Second section of Tab.\ref{Tab:exps} lists some of the elements we have deemed capital to satisfactory collect reliable data.
PUEs should be designed favouring scalability in the sense of allowing the increase of the population sample size or the repetition of the same experiment in another space easily. To make this possible, the experiments must rely on solid and well tested infrastructures, with an appealing user experience to avoid frustration by participants. When considering the experimental setup, we used several strategies to foster participation and ensure the success of the experiments. 

First, we physically set up the PUEs in very particular contexts in urban areas and, in all cases, we placed them in crowded (moderately to highly dense) places to easily reach volunteers. In other words, we preferred to go where citizens were instead of encouraging them to come to our labs. To make this possible the City Council has for instance offered specific windows in a couple of festivals as hosting events (Bee-Path(1), Bee-Path(2), Cooperation(1), Mr. Banks and Dr. Brain experiments). This meant that we had to adapt to these out-of-the-lab and in-vivo specific contexts, the logistics and the composition of research teams thus unavoidably becoming more diverse, complex, heterodox and highly multidisciplinary. In collaboration with the event organizers, we then habilitated a specific space of reduced dimensions for the experiment, where the volunteers (their typology was different in each case) could participate through a recording device. 

Second, PUEs demands that devices used by participants to collect and manipulate data, either in a active or passive way, must be familiar to them. In our experiments we have designed specific software running in laptops (Cooperation(1) and Cooperation(2) experiments), mobile phones (Bee-Path(1) and Bee-Path(2) experiments) and tablets (Mr. Banks and Dr. Brain experiments). However, it may also be possible to use video cameras, photo cameras, or any other sophisticated device as long as participants easily get familiarized with it after few instructions or a tutorial. This sort of infrastructure is in the end what allows to carry out experiments in a participant's quotidian (not strange or in-vivo) environment. Initially, we overlooked this aspect in the set-up design and allocation of resources, but having friendly user interface is important to allow people to behave \emph{normally}. Similarly, both instructions and interface should be understandable and manageable for people of all ages. 

Third, in order to study social behaviour in different environments, the experiments need to be adaptable, tunable, transportable, versatile and easy to set up in different places. All devices mentioned before fulfil this requirement as well. 

Fourth, PUE are typically one-shot since we are hosted in a festival, a fair or a class-room, which means concentrated in time (duration from 0.5 to 2 days) and no chance for a second shot. All collectable data could be threatened if something goes wrong. A strong beta testing and defensive programming is imperative to ensure that collected data is certainly reliable. It is also necessary to anticipate potential problems: One must be flexible enough to be able to retrace alternative research questions on the fly if PUEs location and conditions are not fully satisfactory to respond to the initial research purposes.

Finally, numbers matter and experiments must reach enough statistics to perform rigorous analysis and this needs to be carefully taken into account during the design phase of the experiment (see Fig. \ref{fig:01}). Typically, the more expensive devices are, the less number of data collectors you can have so the capacity to collect data is affected, and this is not very effective strategy in a rather short time one-shot event. Therefore, cheap infrastructures favour scalability at the end. Alternatively new collaborators needs to be found or an extra effort is required for finding a sponsorship (for instance related to science outreach) which in any case will make the experiment preparation phase more complicated. Scalability is interesting indeed but it has its side-effects as well: Relying on user-provided infrastructures such as smart phones can influence a lot the quality of the data and its normalization (one of the central problems related to the Big Data paradigm). In the Bee-Path(1) experiment, where we have used the GPS of participants own smartphones, the cleaning process has been far more laborious than in the other cases due to this circumstance (see numbers in Tab. \ref{Tab:exps}). In contrast, Cooperation(1), Cooperation(2), Mr. Banks and Dr. Brain experiments, where we designed and programmed the software of the participants experience, have not required much posterior treatment.

\subsection{Public Engagement tools and strategies}
 
The PUEs are physically based and rooted on particular, temporal and local contexts. This delimited framework allows to concentrate dissemination resources and efforts on a given spot over a certain time, which increases the effectiveness and efficiency of the campaign in terms of workforce and budget. Additionally, the one-shot nature of PUEs allows us to avoid the problem of keeping participants engaged in an activity spanning long periods of time, but consequently relies completely on constantly renewing the base of participants (which may require higher dissemination). To this end, the initial action was the creation of a census of volunteers shared by all research group members of the Barcelona Citizen Science Office. 

Another factor of capital importance relates to the contact between researchers, organizers and citizens in the set-up of the experiment. This allows for a nice dialogue and exchange of views that in turn helps in framing the scientific question being studied, as well as developing possible improvements for future experiences. This can be done by stimulating participants' curiosity on the experiment and research associated to it. The research question should be focussed and understandable for anyone who is not expert on the field.

To engage in such a dialogue, however, requires certain steps. Firstly, one needs to attract potential participants with an appealing set-up. This includes location in the physical space, but also an effective diffusion campaign on the days prior to the experiment (see preparatory actions in Tab. \ref{Tab:exps}). It is important to offer a harmonized design with common themes that citizens can relate to the experiment. To make our material appealing, we collaborated with an artist (Bee-Path(1) and Bee-Path(2) experiments) and a graphic designer (Cooperation(1), Cooperation(2), Mr. Banks(1), Mr.Banks(2) and Dr. Brain experiments) whose main contribution was the creation of characters associated to each experiment. The function of these characters, close to the world of cartoons, was to attract the public attention but also to present the experiment as an attractive game since one of the most powerful elements to engage people in an activity is the expectation to have fun. It is certainly possible to maintain scientific rigour while using \emph{gamification} strategies to create a playable atmosphere for the study, thus transforming it into a more complete experience \citep{Morris:2013}. Moreover, actors were used as human representations of these characters (Mr. Banks and Dr. Brain). The actors were indeed one important element to bring on-site attractiveness to the experiment, along with: a large team of scientists and facilitators present (up to a rotating team of 10 people), an optimum and visible location inside the event space and material/devices to promote the experiment such as screens to visualize on real-time the results of the experiments or promotional material (flyers and merchandising). Based on the experience, we have optimized all these ingredients and we have even included some elements of scenography in the last experiment done within the Barcelona DAU Board Games Festival (Dr. Brain).

But getting enough people is not enough. Additionally, we must also seek for universality in our population sample. The experiment must be designed in a way that people from all ages and conditions can really participate. Besides, a PUE has to be transportable with a minimal cost once it has been implemented one time in order to be reproduced in different environments (which may favour certain types of population). As an example case, Cooperation(1) and Cooperation(2) experiments are very illustrative. In Cooperation(1) experiment, we found how different age groups, specially children ranging from 10 to 16 years old, behaved in a different way and cooperated with different probability. Apparently children were more volatile and less prone to cooperate than the control group in a repeated Prisoner Dilemma. Fifteen months later we repeated the experiment in a Secondary school (12-13 years old, Cooperation(2) experiment). On one side, the results in this case showed the same levels of cooperation than the control group in Cooperation(1) experiment. However, on the other side, same volatile behaviour, more than Cooperation(1)'s control group, was again found. Therefore, thanks to the repetition of the experiment we rejected the early idea of different cooperation levels in children, but at the same time it strengthened the children’s volatile behaviour claim.


\subsection{Outcome and Return to the Public}

Last but not least come factors related to the management of the aftermath of the experiment (fourth section of Tab. \ref{Tab:exps}). The PUEs as we have implemented them are intrinsically cross-disciplinary and involve a large number of agents and institutions which in turn may have diverging interests and expectations regarding the outcome of a particular experiment. Any successful PUE must be able to accommodate all these interests and create positive collaboration environments where all actors contribute in a conformable way.

Organisers of a festival will for instance find in the PUEs an innovative format with participatory activities to add to their program (Cooperation(1), Mr. Banks and Dr. Brain experiments). PUEs also can be a transparent and pro-active data gathering system to provide information and analysis about the event itself, useful for its planning and improvement. The Bee-Path(1) experiment studied how visitors moved around a given space and obtained fact-based arguments to improve the spatial distribution in future editions of the fair where it was developed. The city and local administration will find in PUEs an innovative way to establish a direct contact with their citizens, to co-create new knowledge valid for the city interests and eventually to generate a census of highly motivated citizens ready to collaborate in these kind of activities. Scientists will obviously try to publish a new research paper with the data gathered.
 
All these expectations are very different and should organically converge to successfully run a collective experiment \citep{latour:2006}. However, we should also not forget to carefully include the expected return of our central actors: the dedicated volunteers. Their contribution is essential in CS practices~\citep{irwin:1995} and it is therefore completely fair to argue that citizens accepting to participate in these initiatives need to clearly see a benefit from their perspective comparable (albeit different) to that of City Council, festival organisers, scientists or any other contributor. Moreover, the high concentration in space and time jointly with the intense public exposure of the PUEs increase even more the volunteer's expectations compared to other ordinary cases in CS~\citep{Kim:2014,lintott:2008}. The face-to-face relationship established between researchers and citizens in all the PUEs we have been running certifies that this is a very delicate issue that needs to be managed with real care.

Any PUE should manage expectations at three different time scales: Short (during and right after the experiment), medium (a week or a month after), and long times (months or even a year later). In some of the experiments we have failed in this purpose in at least one of the three time scales since we did not properly anticipate the strong effort required to respond to volunteer's expectations. 

Short time scale responds to a basic curiosity. This point is related to that of engagement and experiment set-up: The physical presence of scientists (with no mediation) allows to explain the experiment and motivate people. Also the introduction of large screens where the progress of other participants can be followed on real-time helped in this matter. In the Bee-Path experiments (Bee-Path(1) and Bee-Path(2)) we showed the GPS locations of the participants on a map, and in the Mr. Banks experiment we were giving a ranking list of the best players (best performances among the participants). This information was intended to reinforce participation and it was also chosen in a way that it was minimally distorting the questions addressed and thus not influencing the results of each experiment. 
Medium time scale relates to expectations about the results of the experiments. Participants want to know whether the set-up was successful and whether they performed \emph{well enough}. In order to complement their experience, it is important to keep them informed about the outcomes of the PUE. An example of a medium timescale is the case of the Dr. Brain where a personalized report of the performance during the experiment was sent to each participant by e-mail. In some cases and based on these results, this generated as well new conversations between scientists and citizens.
The last time scale to manage is a more formal way of presenting the results of the study through public presentations and talks. This was the case for the Bee-Path(1) experiment, where public conferences and even a summer course dedicated to (graduated and undergraduated) students interested in CS practices was done. 

All the stages being discussed are important to the success of the PUEs and should be clearly exposed to volunteers even before they agree to participate. The return to volunteers in all these scales is a key ingredient to build a critical mass of engaged citizens, not only for further experiments but also to fulfil the non-strictly scientific objectives of the work. Being a scientist, the direct relation with volunteers helps to improve the message and the way of delivering the message, to refine the understanding of the phenomena of your experiment and even to upgrade a given experiment in future venues. A final positive side effect of this contact is also the raise of public awareness towards the difficulty and importance of science. As it can be seen, growing the project around a rich and functional webpage helps in harmonizing all the time scales discussed and opens new and interesting perspectives to bridge PUEs with other CS on-line based practices. It also serves as an efficient way to communicate results and news of the project to interested citizens and it could be used to improve data handling and sharing standards by allowing participants a direct access and management of their personal recordings.

\section{Discussion}
\label{Discussion}

The advent of globalization and the fast track taken by innovation \citep{chesbrough:2006}, combined with enormous challenges demand for answers at a very fast pace. Deeply interwined global and local actions are necessary to face societal challenges such as the continuous growth of humanity, the effects of climate change or even the needs for collective decision-taking mechanisms ready for an effective policy-making. These urgencies collide with the typically long process of scientific research, and is affecting the philosophy, the resources and the methods behind the scientific method \citep{franzoni:2014}. Society expects much from us being scientists but still lacks of knowledge and reflection about the ways of a more collaborative, public, open and responsive research. Citizen Science practices, even if they may not provide definitive answers to societal challenges, expects to close the gap between public and researchers or in the worst case scenario to increase social awareness on the treated problems. CS practices can also allow science to furnish ready made solutions in public and with the active role of citizens. 

In this work we have presented our experience in bridging Computational Social Sciences with Citizen Science philosophy, which in our case has taken the form of what we call Pop-Up Experiments (PUEs). We hope that this work serves as guideline to groups willing to adopt and expand such practices and that it opens the debate about the possibilities (but also the limitations) these approaches can offer. The flexibility of Behavioural Experiments can be combined with the strengths of Big Data to create a new tool capable to generate a new collective knowledge. We have conceptually identified the main issues to take into account whenever planning a CS research into CSS field. We have grouped the challenges into three categories according to our experience: Engagement, Infrastructure and Return. Furthermore, we have explained the solutions we implemented under the framework of a PUE, providing practical examples grounded in our own experience. The importance of team work and of widening the scope to consider questions not directly related with lab work has been highlighted as well as the need to work hand in hand with both public and other social actors. Other technical aspects of the approach have also been reviewed.

We have supported the idea of abandoning the \emph{Ivory Tower} and opening up science and its research processes. Indeed, CS research essentially relies in the citizenship collaboration, but not just in the passive data gathering. We think that to place Distributed Intelligence (with contributions from experts and amateurs) in the very core of scientific analysis can be also a valid strategy for obtaining rigorous and valuable results. We also believe that the PUEs we herein present can potentially empower citizens to take their own civic actions relying on a collectively constructed facts-based approach \citep{mcquillan:2014}. To co-create and co-design with citizens a \emph{smart city} along the Big Data paradigm will then be much easier and even more natural since we will share interests and concerns during the whole research process. Data gathered in the wild or in-vivo contexts could be thus understood as truly public and open while data ownership and knowledge would be shared from the very start \citep{callon:2003}. Our future venues and experiences will take deeper inspiration from Open Source, do-it-yourself, do-it-together and makers movements \citep{hatch:2014} which facilitates learning-by-doing, low-cost and heuristic skills to everybody. A fresh look to problems can contribute with innovative and imaginative ideas that at the end can reach \emph{out of the box} solutions. However, being scientists, we will also have to find the way to reconcile unorthodox and intuitive forms with the standards and methodologies of the world of science. Lessons shall be learnt from the Open Prototyping approach where an industrial product such as a car can be shaped by an iterative process where the company owning the product has no problem to allow for inputs coming from the outside \citep{bullinger:2011}. Some other clues can be found in the form of collective experimentation where a fruitful dialogue can be stablished among the \emph{matters of concern} raised by citizens and the \emph{matters of fact} raised by scientists. \cite{latour:2009} already introduced these concepts and discussed the symbiotic relationship by taking the case study of Ecologism (a civic movement) and Ecology (a scientific activity). There are still many aspects to test and explore following this approach within the CSS research field.

We would also like however to shortly present open questions related to the way we perform science nowadays, and echo fundamental contradictions that science is not properly handling in this globalization era \citep{franzoni:2014}. Citizen practices yield nice outputs for the communities, yet they require a lot of effort by scientists, with the downside of providing very low (formal and bureaucratic) professional recognition. Open social experiments demand a high level of involvement in cooperation with non-scientiffic actors which may divert professional researchers from the activity for which they are evaluated: publication of results. Furthermore, such experiments often involve multidisciplinar teams, which then may find difficulties finding the appropriate journals to publish, facing difficult acceptance in stablished communities. We thus urge scientific community to actively recognize the valuable advantages of performing science with our proposed experimental framework.

Lessons learned must be shared, both within the community (like this paper is trying to do) and outside in public spaces including also public institutions and policy makers. Science and Citizen Science are mostly publicly funded and therefore belong to society. Internet provided new ways in which new relations among science and society can be strongly reinforced. We believe that this is good for everyone by raising Science concern, by enhancing participation and, most importantly, by exploring new effective ways to push knowledge further. We hope that the present work helps in theoretically establishing the concept of Pop-Up Experiments and encouraging the adoption of CS practices in science, in whichever the field.

\section{Material \& Methods}\label{sec:methods}

In this section we provide description of our experience throughout four years of experiments in performing pop-up events using CS practices in the city of Barcelona. Here we detail the experimental methods and briefly summarize the outcome. All experiments have been performed in accordance with institutional (from the University of Barcelona in all cases except for Cooperation(1) experiment which was from Universidad Carlos III) and national guidelines and regulations in data privacy (following the Spanish law LOPD). All interfaces being used included an informed consent from all subjects. Data collected was properly anonymized and unrelated from personal details which in our case was age range, genre, level of studies and electronic address (e-mail).

\subsection{Bee-Path(1) experiment}

The aim of this experiment is to study the movement during the exploration of an outdoor science and technology fair where several stands with activities were located in an area of approximately 3 ha inside a public green park. The experiment took place during the weekend of 16th and 17th of June of 2012, specifically in Saturday afternoon (from 16h to 20h) and the morning of Sunday (from 11h to 15h). The participants had very different interests, origins, background and ages and organisation of the event estimated that 10,000 people visited the fair. The Bee-Path information stand was located at the main entrance where visitors were encouraged to participate in the experiment by downloading an App from their mobile phones. After a very simple registration and instructions to activate the App, participants were left wandering around the fair while being tracked.
After a laborious cleaning process we could analyse the movement and trajectories of 27 subjects among those records provided by 101 volunteers. We found spatio-temporal patterns in the movement and we developed a theoretical model based in Langevin-dynamics in a gravitational potential landscape created by the stands. This model based was able to explain the results of the experiment and to model different scenarios with other spatial configurations of the stands. Scientific paper is been submitted for publication \citep{Gutierrez:2015}. The project description, results and data are freely accessible on the webpage \url{www.bee-path.net}.

\subsection{Bee-path(2) experiment}

Following the previous Bee-Path(1) experiment this setup was also focussed in studying the movement, but in this case we were interested in human searching patterns or how people move when explores a landscape to find something. The experiment took place in the next edition of the same science and technology fair (June 15th and 16th of 2013). Participants also downloaded a similar App that were tracking them, but in this case were instructed to find 10 mannequins hidden in the park.
Numerous problems were encountered due to several technological and non-technological factors which impeded a satisfactory performance of the experiment. Technological troubles were principally two; The low accuracy of recordings due to the proximity of regional parliament where the wi-fi and mobile phone coverage is inhibited, and the limitation of the App to perform adequately when running in low-end devices. Non-technological part includes a bad placement (far from the entrance) of our stand where the recruitment of volunteers was hard and the unexpected fact that some vandals changed, stole and even manipulated the mannequins' position. Notwithstanding the failure, we learnt important lessons from these complications.

\subsection{Cooperation(1) experiment}

Here we explored how important is the age in the emergence of cooperation when people are facing repeated Prisoner Dilemma. The experiment was carried out with 168 volunteers selected from the attendants to DAU Barcelona Festival 2012 (1st Board Game Fair of Barcelona, December 15th and 16th). This volunteers' set was divided into 42 subsets of 4 players according to the age, seven different age groups plus one control group where subjects where not distinguished by age. Each subset made up a game where the four participants played 25 rounds (although they were not aware of it) deciding between two colors associated to a certain prisoner's dilemma payoff matrix. Participants played a 2x2 PD game with each of their 3 neighbors, choosing the same action for all opponents. In order to play with an incentive, they were remunerated with real money proportional to the final score. During the game volunteers interacted through software specially programmed for the experiment and installed in a laptop. They were not allowed to talk or signal in any way, but to further guarantee that potential interactions among players would not influence the results of the experiment, the assignment of players to the different computers of the room was completely random.
In this experiment, together with the Jesu\"ites Casp described in the next subsection, we found that the elderly cooperate more and there exist a behavioural transition from reciprocal, but more volatile behaviour to more persistent actions towards the end of adolescence. For further details consult \cite{gutierrez:2014}.

\subsection{Cooperation(2) experiment}

The purpose to repeat the Cooperation(1) experiment at DAU festival was to confirm the apparent tendency of children to cooperate less than the averaged population. Thus, we repeated the experiment just to augment the pool of subjects from this age range what allowed us to be more statistically accurate. We analysed the performance of 52 students of secondary school (Jesu\"ites Casp experiment) ranging from 12 to 13 years old. The methods and protocols were the same as in Cooperation(1) experiment, as well as the software installed in its own laptops.
The results of this experiment rejected the hypothesis that children cooperated less in average but confirmed at the same time their more volatile behaviour, as it describes \cite{gutierrez:2014}.

\subsection{Mr.Banks(1) experiment}

Mr. Banks experiment was set up to study how non-expert people make decisions in uncertain environments, concretely assessing their performance when trying to guess if a real financial market price would go up or down. We analysed the performance of 283 volunteers (from the approximately 6,000 attendants) to DAU Barcelona Festival 2013 on December 14th and 15th. All the volunteers played via an interface specifically created for the experiment that was accessible through identical tablets only available in a specific room under researcher's surveillance. Devices showed in the main screen the historic of daily market price curve and some other information like 5-day and 30-day average windows curves, high-frequency price of previous day, the opinion of an expert, price direction of previous days and price directions of other markets around the world. All price curves and information was extracted from real historical series. Participants could play in four different scenarios with different time and information availability constrains. In each scenario they were required to make guesses during 25 rounds while every click in the screen was recorded. Each player started with 1,000 coins and he/she earned 5\% over their current score if the guess was correct or got a negative return of the same size if he/she was wrong. We used gammification strategies and we did not provide any economic incentive in contrast with Social Dilemma experiments.
The analysis of 18,436 recorded decisions and 44,703 clicks allowed us to claim that participants tend to follow intuitive strategies called Market Imitation and Win-Stays Lose-Switch. These strategies are less followed when there is more time to make a decision or some information is provided \citep{Gutierrez:2015b}. Either the experiment and information of the project is available at \url{www.mr-banks.net}.

\subsection{Mr.Banks(2) experiment}

Here we repeated the Mr.Banks(1) experiment in a different context in order to study the reproducibility of the results obtained. The interface and the experimental set-up were the same as in Mr.Banks(2) but the typology of participants and the type of event were significantly different. The experiment, named “Hack your Brain” in the conference programme, was situated at the main entrance of the CAPS2015 conference, the International Event on Collective Awareness Platforms for Sustainability and Social Innovation, held in Brussels (July 7th and 8th of 2015). The 42 volunteers that played the game provided 2,372 recorded decisions. The volunteers were all registered participants of the conference with very diverse profiles (scientists mostly from social sciences, social innovators, designers, social entrepreneurs, policy-makers, etc.). The results showed that the results of the Mr.Banks Experiment had a good reproducibility as the percentage of correct guesses were similar in Mr.Banks(1) and Mr.Banks(2) (53,43 \% and 52,74 \%, resp.) as well as the percentage of market-up decisions (60,83 \% and 60,48 \%, respectively). A deeper analysis of the results is underway to check if the strategies used by Mr.Banks(1) volunteers are the same than those used by Mr.Banks(2) volunteers.

\subsection{Dr.Brain experiment}

In this work, we report the results of a lab-in-the-field experiment that allows for a phenotypic characterization of individuals when facing different social dilemmas. Instead of playing with the same fixed payoff matrix as in Cooperation experiment, here the values and the neighbors changed every round. We discretized the $(T,S)$-plane as a lattice of $11 \times 11$ sites allowing us to explore up to 121 different games grouped in 4 categories: Harmony Game, Stag Hunt Game, Snowdrift Game and Prisoner Dilemma Game. Each player was given a tablet with the application of the experiment. The participants were shown a brief tutorial, but were not instructed in any particular way nor with any particular goal in mind. They were informed that they had to make decisions in different conditions and against different opponents in every round. Due to practical limitations, we could only host around 25 players simultaneously, so the experiment was conducted in several sessions over a period of two days. In every session, all individuals played a different, randomly picked number of rounds between 13 and 18. The total number of participants in our experiment was $541$, adding up to a total of $8,366$ game actions collected. In order to play with an incentive, they received back coupons for a prize of 50 euros to spend in stores of the neighborhood. During the game volunteers interacted through software specially programmed for the experiment and installed in a laptop. They were not allowed to talk or signal in any way, and again were placed spatially at random.
This experiment concludes that we can distinguish, empirically and without making any assumption, five different player's behaviour or phenotypes not theoretically predicted. Paper is been submitted for publication \citep{Poncela:2015}.

\section*{Disclosure/Conflict-of-Interest Statement}
The authors declare that the research was conducted in the absence of any commercial or financial relationships that could be construed as a potential conflict of interest.

\section*{Author Contributions}

OS, MGR, IB, and JP have equally conceived and written the work. All authors have approved the final version of the manuscript.

\section*{Acknowledgments}
We would like to acknowledge the participation of at least 1,255 anonymous volunteers who have
made this research possible. We specially want to thank 
Mar Canet, Nadala Fern\'andez, Oscar Mar\'in from Outliers, Carlota Segura, Cl\`audia Payrat\'o, Pedro Lorente, Fran Iglesias, David Rold\'an, Marc Angl\'es, and Berta Paco for all the logistics and full collaboration to make the experiments possible in one way or another. We also acknowledge co-authors Anxo S\'anchez, Yamir Moreno, Carlos Gracia-L\'azaro, Jordi Duch, Juli\'an Vicens, Julia Poncela-Casasnovas, Jes\'us G\'omez-Gardenes, Albert D\'iaz-Guilera, Federic Bartomeus, Aitana Oltra, and John Palmer in the subsequent research articles from the experiments herein reported. We also thank to the director of the DAU (Oriol Comas) for giving us the opportunity to make the three experiments in DAU Barcelona Festival. Last but not least, we are greatly indebted to Barcelona Lab programme promoted by the Direction of
Creativity and Innovation from the Barcelona City Council led by In\'es Garriga for their
help and support for setting up the experiments at the Barcelona Science Fair (Parc de la Ciutadella, Barcelona public park) and at the DAU Barcelona Festival (Fabra i Coats, Creativity Fabrique of the City Council).
 
\paragraph{Funding\textcolon} This work was supported in part by Barcelona City Council (Spain), RecerCaixa (Spain) through grant {\it Citizen Science: Research and Education}, by MINECO (Spain) through grants FIS2013-47532-C3-3-P and FIS2012-38266-C02-02, by Generalitat de Catalunya (Spain) through grants 2014-SGR-608 and 2012-ACDC-00066, and by Fundaci\'on Espanyola para la Ciencia y la Tecnolog\'ia (FECYT, Spain) through the Barcelona
Citizen Science Office project of the Barcelona Lab programme.

\bibliographystyle{frontiersinSCNS} 
\bibliography{Bibliography}


\end{document}